\documentclass[12pt,preprint]{aastex}

\newcommand{\f}[2]{\frac{#1}{#2}}

\shorttitle{Modeling GRB X-ray flares}
\shortauthors{Maxham \& Zhang}

\begin{document}

\title{Modeling Gamma-Ray Burst X-Ray Flares within the Internal Shock Model}

\author{Amanda Maxham and Bing Zhang}
\affil{Department of Physics and Astronomy, University of Nevada,
  Las Vegas}

\begin{abstract}
X-ray afterglow light curves have been collected for over 400 Swift
gamma-ray bursts with nearly half of them having X-ray flares
superimposed on the regular afterglow decay.
Evidence
suggests that gamma-ray prompt emission and X-ray flares share a
common origin and that at least some flares can only be explained by
long-lasting central engine activity.  We have developed a shell model
code to address the question of how X-ray flares are produced within
the framework of the internal shock model. The shell model creates
randomized GRB explosions from a central engine with multiple shells
and follows those shells as they collide, merge and spread, producing
prompt emission and X-ray flares. We pay special attention to the time
history of central engine activity, internal shocks, and observed
flares, but do not calculate the shock dynamics and radiation
processes in detail.  Using the empirical $E_p-E_{iso}$ (Amati)
relation with an assumed Band function spectrum for each collision and
an empirical flare temporal profile, we calculate the gamma-ray
(Swift/BAT band) and X-ray (Swift/XRT band) lightcurves for arbitrary
central engine activity and compare the model results with the
observational data.  We show that the observed X-ray flare
phenomenology can be explained within the internal shock model.
The number, width and occurring time of flares are then used to
diagnose the central engine activity, putting constraints on the
energy, ejection time, width and number of ejected shells.
We find that the observed X-ray flare time history generally
reflects the time history of the central engine, which reactivates
multiple times after the prompt emission phase with progressively
reduced energy.
The same shell model predicts an external shock X-ray afterglow
component, which has a shallow decay phase due to the initial
pile-up of shells onto the blast wave. However, the predicted
X-ray afterglow is too bright as compared with the observed
flux level, unless $\epsilon_e$ is as low as $10^{-3}$.
\end{abstract}


\section{Introduction }
The study of gamma-ray bursts (GRBs) has
been greatly advanced following the launch of the Swift
Gamma-Ray Explorer (Gehrels et al. 2004) on November 20, 2004.
Thanks to its rapid slewing capability, multi-wavelength
observations of GRB afterglows, usually as soon as $<100$
seconds after the burst trigger, have been performed
regularly for many bursts. The X-ray telescope (XRT, Burrows
et al. 2005a) aboard this satellite has given us unprecedented
access to early afterglows in the X-ray band. For most GRBs,
XRT has observed smoothly decaying power-law or broken power-law
afterglows (Nousek et al. 2006; O'Brien et al. 2006; Zhang et al. 2007;
Liang et al. 2007, 2008; Evans et al. 2009).  In about half of GRBs,
superimposed on the background X-ray afterglow, one or more X-ray
flares are seen (Burrows et al. 2005b; Romano et al. 2006; Falcone
et al. 2006, 2007; Chincarini et al. 2007).

Flares among the bursts share some common properties, which include
the following:

\begin{itemize}

\item The morphology of flares is similar: they all show a smooth,
rapid rise and a rapid fall (Romano et al. 2006).

\item They are superimposed on a background power-law decay
afterglow component with the slope before the flaring equal to that
after the flare (Burrows et al. 2005c).

\item The width of flares is typically narrow, with
$\delta t / t \sim 1/10$ on average, where $t$ is the emission time
of the flare.  However, flares can become wider ($\delta t / t$ becomes
larger) as the time of their emission increases, and no sharp flares
are seen at late times (Chincarini et al. 2007; Kocevski et al. 2007).

\item Similar flaring activity has been detected in both types of GRBs:
those believed to be of massive star origin (Type II, typically long
duration) and those believed to be of compact star merger origin
(Type I, typically short, e.g. in GRB 050724, Barthelmy et al. 2005)
\footnote{For a full discussion of the two physically distinct
types of GRBs, see Zhang et al. (2009).}.
This suggests that the
phenomenon is insensitive to the progenitor type (Perna et
al. 2006).

\item Flares are spectrally harder than the underlying afterglow
(Burrows et al. 2005b; Romano et al. 2006; Falcone et al. 2006).

\item For a small sample of bright flares whose time-dependent
spectral analysis can be performed, flares are found to soften
as they decay (Burrows et al. 2005b,c; Falcone et al. 2006, 2007),
reminiscent of GRB prompt emission.

\item Average X-ray flare luminosity decays with time as a power-law with
slope $\sim$ -1.5 for a sample of flares from many GRB's.  This temporal
relationship also seems to hold among flares from a single
multi-flared burst (Lazzati, Perna \& Begelman 2008).
\end{itemize}

Flares also vary from burst to burst, from flare to flare, in the
following ways:
\begin{itemize}

\item The shape of flares can be best fit with a number of different
profiles: gaussian, log-normal distributions, exponential or power-law
rise or decay with differing slopes for both the rising and falling
portion of the flare (Chincarini et al. 2007).

\item The number of flares seen per burst varies from modal value of
1, mean value of $\sim$ 2.5 and maximum value of 8
(Chincarini et al. 2007).

\item The fluence seen in flares can vary from a few percent of the
burst fluence to larger than the burst itself (e.g. for GRB050502B,
Burrows et al. 2005b,c; Falcone et al. 2006).

\item The emission time of flares varies from perhaps before the
slew of XRT (e.g. GRB060607) to late flares at $10^4$ seconds
(e.g. GRB050904, Cusumano et al. 2006) to $10^5$ seconds (for the
Type I GRB050724, Barthelmy et al. 2005). Some flares are
superimposed on other flares (e.g. GRB050916, Burrows et al. 2005b,c;
Chincarini et al. 2007). Although most flares happen early, from 100
to 1000 seconds, the distribution of $t$ tails off to $10^6$ seconds
(Chincarini et al. 2007).
\end{itemize}

A few properties of X-ray flares suggest that they are connected to
internal emission processes due to a restarting of
the central engine at late
times (Burrows et al. 2005b; Falcone et al. 2006; Romano et al. 2006;
Zhang 2007). The arguments in favor of such a ``late internal''
model (in contrast to the external shock model) are the following:
First, the rapid rise and fall of flares with $\delta t / t
\sim 1/10$ strongly disfavors external shock models that involve a
large angular area over which radiation would be emitted (Zhang et
al. 2006, see also Ioka et al. 2005; Fan \& Wei 2005; Lazzati \&
Perna 2007).  Second, the connected, underlying continuum which has
the same slope both before and after the flare suggests that flares
are not related to the canonical afterglow and are instead
superimposed on top of this regular afterglow decay (Chincarini et
al. 2007).  Third, given the same observed X-ray flare amplitude,
the internal model requires a much smaller energy budget than the
external shock model (Zhang et al. 2006). While the internal model can
produce significant X-ray flares with energy much less than that during
the prompt emission, the external shock model requires an energy
budget at least comparable to that of the initial blast wave in
order to make a noticeable change in flux (Zhang \& M{\'e}sz{\'a}ros,
2002). Internal models are much more ``economical'' as far
as energy budget is concerned. Next, Liang et al. (2006) have shown
that as long as the central engine clock is reset to zero at the
beginning of each flaring episode, the curvature effect can naturally
explain the spectral index and the temporal decay index of the
decay phase of flares. Finally, although there is no correlation
between the number of prompt emission burst pulses and X-ray flares
in any given burst (Chincarini et al. 2007), X-ray flares do exhibit
the same spectral softening as GRB prompt emission (Burrows et al.
2005b,c; Falcone et al. 2006, 2007). This suggests that the properties
of X-ray flares make them likely to be caused by the same mechanism
as prompt emission seen in gamma-rays.

Any X-ray flare model must
be able to explain the bulk similarities and differences among flares
enumerated above. The leading ``internal'' model is the internal
shock model (Rees \& M\'esz\'aros 1994), which invokes internal
collisions of shells within an unsteady central engine wind.
Previous internal shock models have focused on interpreting prompt
gamma-ray emission properties (e.g. Kobayashi et al. 1997;
Panaitescu et al. 1999; Spada et al. 2000; Guetta et al. 2001).
Kobayashi et al. (1997) have shown that internal shocks can explain
the highly variable profiles seen in GRBs.   In their model, shells
of matter with varying Lorentz factors ejected from the central engine
produce about as many collisions as the number of shells, and the time
of ejection of the shell is highly correlated to the time of collision
of the shells, indicating that there is essentially no delay between
energy ejection by central engine emission and the time when that
emission is seen. Recently, using a model where collisions
are not perfectly inelastic, Li \& Waxman (2008) studied the effect
of residual collisions in optical emission which would be
slightly delayed from gamma-ray prompt emission.

Within the context of X-ray flares, the internal shock model has
been discussed by a number of authors. Zhang et al. (2006) and
Fan \& Wei (2005) discussed how a late internal shock may produce
a softer flare than prompt gamma-ray emission. Wu et al. (2005)
discussed both late internal and external shock models to study
X-ray flares and concluded that at least some flares have to be
produced by late internal shocks. Lazzati \& Perna
(2007, see also Zhang 2007) proved the suggestions of Burrows
et al. (2005b) and Zhang et al. (2006) that late flares must require
late injection of shells and cannot be produced by late collisions
of shells ejected during the prompt phase. Yu \& Dai (2009) studied
the shock physics and radiation processes of late internal shocks
that may be responsible for X-ray flares.

In this paper, we focus on another aspect of the internal shock
model for X-ray flares. Extending the work of Kobayashi et al. (1997)
to concentrate on X-ray flares, we have created a GRB fireball model
with a central engine that can eject multiple episodes of matter
shells with any Lorentz factor, thickness and mass distributions.
This code is used to address the question of what kind of central
engine activities are demanded in order to reproduce the observed
properties of X-ray flares. The conclusion drawn from this study
may be taken as the requirements of some central engine models
on X-ray flares (e.g. King et al. 2005; Perna et al. 2006; Proga
\& Zhang 2006; Dai et al. 2006; Lee et al. 2009)\footnote{Other
internal dissipation models for X-ray flares
(e.g. Panaitescu 2008) may not subject to these requirements.}.
In \S2, we first examine the physics of two-shell collisions
and describe our phenomenological treatment of the spectrum and
lightcurve of individual collisions. In \S3, we discuss
blast wave evolution, which sets an outer boundary for the internal
collisions that are relevant to X-ray flares. We then dedicate
our discussion (\S4) to multiple collisions for multiple ejection
episodes and use the observations to diagnose the required
central engine activities. Our results are summarized in \S5
with some discussion.

\section{Two-Shell Interaction}
Our model represents the GRB central engine as a variable source
that ejects many randomized shells of matter over some period of
time.   Shells ejected later with higher Lorentz factors catch up
with slower shells, colliding and creating ``internal shocks''
(Rees \& M{\'e}sz{\'a}ros, 1994).  As the modeled shells move
outward from the central engine, the energy released from each
collision is calculated, allowing the two colliding shells to
combine and continue to move outward to collide with other shells.
In the model, each shell is given four initial parameters: ejection
time $t_{ej}$, relativistic Lorentz factor $\gamma$, mass $m$,
and thickness $\Delta$.

We assume that shell collisions are inelastic, so that a fast shell
(f) and a slow shell (s) merge to form a merged shell (m). Using
conservation of momentum and energy, the Lorentz factor of the
combined shell can be written
\begin{equation}
\gamma_{m} \simeq \sqrt{\frac{\gamma_{f} m_{f}+ \gamma_{s} m_{s}}
{\frac{m_{f}}{\gamma_{f}}+\frac{m_{s}}{\gamma_{s}}}}.
\label{momentum}
\end{equation}
Each collision then releases an internal energy given by (Kobayashi
et al. 1997)
\begin{equation}
E_{int}=(\gamma_{f} - \gamma_{m})m_{f} c^{2} + (\gamma_{s} -
\gamma_{m}) m_{s} c^{2}.
\end{equation}
The efficiency of each collision can be defined as
\begin{equation}
\eta = \frac{E_{int}}{(\gamma_s m_s c^{2} + \gamma_f m_f c^{2})}~.
\end{equation}
It has been known that this efficiency is usually low (Kumar 1999;
Panaitescu et al. 1999) although depending on input parameters, it can vary
in a wide range. To test this, we simulated 100 shells that are
randomly injected during 0-100 s. We allow the mass of each shell
to be randomly drawn in the range of $10^{29}-10^{31}$ g in log space,
and investigate how Lorentz factor contrast affects the distribution
of the efficiency. We randomly generate shell Lorentz factors in
log space within a range of $(\gamma_{min}, \gamma_{max})$. Table 1
shows the mean efficiency of $\eta$ and its standard deviation
$\sigma$ for varying
$\gamma_{max}/\gamma_{min}$. It is evident that the mean value of $\eta$
rises steadily with $\gamma_{max}$/$\gamma_{min}$. However, even for
a very high Lorentz factor contrast $\gamma_{max}$/$\gamma_{min}$ =
1000,  the mean efficiency still only reaches a level of $\sim$ 26\%.
Values of $\sigma$ are on the order of $\eta$, which means that
very low and very high $\eta$ can be expected for extreme parameters
of the two shells.  The right pane of Table 1 shows the efficiency for
the same simulation, but with shell masses standardized to all be $10^{30}$ g.
These results are on order with the efficiency study with varying shell masses.

\begin{table}
\begin{center}
\caption{Efficiency and standard deviation for energy conversion
in the internal shock model.\label{tbl}}
\begin{tabular}{|ccc|ccc|}
\tableline
$\gamma_{max}/\gamma_{min}$ & $\eta$ & $\sigma$ & $\gamma_{max}/\gamma_{min}$ & $\eta$ & $\sigma$\\
\tableline
\multicolumn{3}{|c|}{variable masses} & \multicolumn{3}{c|}{even masses}\\
\tableline
 1000 & 26\% & 26\% & 1000 & 28\% & 27\%\\
 100 & 14\% & 17\%& 100 & 19\% & 20\%\\
 10 & 5\% & 7\% & 10 & 7\% & 8\%\\
 5 & 0.5\% & 0.7\% & 5 & 0.6\% & 0.9\% \\
 \tableline
\end{tabular}
\end{center}
\end{table}

For our model, we do not follow the internal shock physics
in detail. This has been
done by Yu \& Dai (2009), who showed that for reasonable parameters,
the emission from internal shocks can well reproduce the X-ray flare
phenomenology. Our goal is to model many collisions and
to investigate the time history of central engine activity, collision,
and GRB emission. Like many other previous work in this direction
(e.g. Kobayashi et al. 1997; Panaitescu et al. 1999), we adopt an
empirical approach to calculate the spectrum and lightcurve of each
collision event.

{\em Spectral model.} The spectrum of GRBs is typically a
smoothly-joint-broken-power-law spectrum, or ``Band''-function
(Band et al. 1993)
\begin{equation}
N_{E}(E)  = \left\{
\begin{array}{l l}
  A \left( \f{E}{100 {\rm keV}} \right)^{\alpha} \rm{exp} \left(- \f{E}{E_0} \right) & \quad (\alpha - \beta)E_0 \geq E \\
 A \left( \f{(\alpha - \beta)E_0}{100 {\rm keV}} \right)^{\alpha-\beta} \rm{exp} (\beta-\alpha)\left(\f{E}{100 {\rm keV}} \right)^\beta & \quad (\alpha - \beta)E_0 \leq E\\ \end{array} \right. \
\label{Band}
\end{equation}
where $\alpha$ and $\beta$ are the power law photon spectral indices
below and above the break energy $E_0$ in the asymptotic regime,
$A$ is a normalization factor. The break energy can be written as
$E_0 = E_p/(2+\alpha)$, where $E_p$ is the spectral peak energy in
the GRB energy spectrum. Since evidence strongly suggests that GRB
prompt emission and X-ray flares originate from similar physical
events (e.g. Burrows et al. 2005c; Chincarini et al. 2007; Falcone et
al. 2007; Krimm et al. 2007), we assume that the spectrum of X-ray
flares is also a Band-function.

The Band-function parameters of X-ray flares are determined in the
following way in our calculations:
The typical values for the spectral indices are taken as
$\alpha=-1$, and $\beta=-2.3$.
For $E_p$, we calculate it through an empirical relation between the
isotropic emission energy $E_{iso}$ and $E_p$, which is
generally valid among GRBs (Amati et al. 2002; Krimm et al. 2009)
and within a burst (Liang et al. 2004; Ghirlanda et al. 2009).
We assume the validity of this correlation
\begin{equation}
E_{p} =100 ~{\rm keV} \left( \f{E_{iso}}{10^{52}~{\rm erg}}
\right) ^{1/2}
\label{Amati}
\end{equation}
and apply $E_{iso}=E_{int}$ to estimate $E_p$. This is because
electrons in the shock are in the ``fast-cooling'' regime, and
lose their energy rapidly. As long as the electron equipartition
parameter $\epsilon_e$ is close to unity, essentially all the
internal energy can be radiated away. For $\epsilon_e \ll 1$,
this estimate gives an upper limit on the brightness of
X-ray flares. We require that the
spectrum-integrated bolometric energy $\int_{0}^{\infty} E N_E (E) dE$
equals $E_{int}$ and then solve for the normalization factor $A$.
Then reinserting this constant, we can calculate the energy
of each pulse $\int_{E_1}^{E_2} E N_E (E) dE$ within each band
$(E_1, E_2)$, e.g. 15 to 150 keV for BAT, 0.1 to 10 keV for XRT.

{\em Temporal model.} The total energy released in a particular band
for a particular pulse is distributed in time throughout the X-ray
flare temporal profile. Different functions of GRB pulses and X-ray
flares have been adopted in the literature (Kobayashi et al. 1997;
Chincarini et al. 2007), but for the purpose of our study, the shape of
the flare profile is not crucial. For simplicity, we adopt the
following temporal profile
\begin{equation}
L(t)= L(t_p)e^{\f{-(t- t_p)^2}{2 (\delta t)^2}}
\label{PDF}
\end{equation}
where the temporal width $\delta t = \Delta/c$ scales as the
physical width of the shell, $\Delta$.
For a shell with initial width $\Delta_0$ moving outward from the
central engine, spreading will occur after the sound wave travels
across the shell at $R_s\sim \gamma^2 \Delta_0$). So in general,
the shell width can be expressed as (M{\'e}sz{\'a}ros et al.
1993; Kobayashi et al. 1999)
\begin{equation}
\Delta  = \left\{
\begin{array}{l l}
  \Delta_0, & \quad R <R_s \\
  \f{R}{\gamma^2}, &\quad R > R_s .\\ \end{array} \right.\
\label{shellspreading}
\end{equation}

We require that the temporal integral of the profile $\int_0^{\infty}
L(t) dt$ equals the internal energy in the specified energy band (BAT
or XRT). When shells collide, the duration of the X-ray flare pulse is defined
by the width of the faster shell. This is because the reverse shock
is typically the one that dominates the X-ray flare emission
(Yu \& Dai 2009). After the collision, which occurs at $R_{col}$, the width of the
the combined shell is taken to be $R_{col}/\gamma_m^2$, i.e. the
width in the spreading regime. The width of the shell keeps spreading
as $R/\gamma_m^2$ thereafter.

In the central engine frame, the shells collide at $t_{col}\simeq
R_{col}/c$. The relevant observation time is\footnote{Usually this
is called observer frame time. However, the central engine and the
observer is in the same inertial frame (with cosmological time
dilation correction). The difference between the two times is due
to a propagation effect, not Lorentz transformation (Zhang \&
M\'esz\'aros 2004).}
\begin{equation}
t_{\oplus,col}=t_{ej}+\frac{(t_{col}-t_{ej})}{2\gamma^2}~,
\label{time}
\end{equation}
where $t_{ej}$ and $\gamma$ can be taken as the ejection time and
Lorentz factor of either of the two colliding shells.
The peak time of a flare is defined as the observed collision time
plus the observed shock crossing time, which is roughly estimated as
\begin{equation}
t_p\simeq t_{\oplus,col}+\frac{\Delta}{c}.
\end{equation}

\section{Blast Wave Evolution}

In order to define ``internal'' collisions, we need to track
the location of the ``external'' shock.
As the first shell moves outward into the ambient medium, it slows
down due to an external shock mechanism as it sweeps up the ambient medium.
As time goes by, more and more trailing shells collide onto the leading
decelerating shell.  The motion of this decelerating ejecta along with
the medium collected along the way, known as the ``blast wave'', is
governed by the following differential equations (Chiang \& Dermer
1999; Huang et al. 2000):
\begin{equation}
\f{d R}{dt} = \beta c = \f{\sqrt{\gamma^2 - 1}}{\gamma}c,
\label{position}
\end{equation}

\begin{equation}
\f{d \gamma}{dm} = \f{-(\gamma^2 -1)}{M},
\label{gammamass}
\end{equation}

\begin{equation}
\f{dm}{dR} = 4 \pi R^2 \rho.
\label{density}
\end{equation}
Here $t$ is the time in the rest frame of the central engine,
$R$ is the distance from the central engine, $\rho$ is the density of
the ambient medium, $\gamma$ is the Lorentz
factor of the shell, $m$ is the swept-up mass, and $M=M_0+\gamma m$ is
the total mass including internal energy of the blast wave, where
$M_0$ is the initial mass of the ejecta. As a result, one has another
differential equation
\begin{equation}
\f{dM}{dm} = \gamma.
\label{Mandm}
\end{equation}
Equation (\ref{position}) simply states how radius changes as a function
of time for an object moving with constant velocity. Equation
(\ref{gammamass}) is a statement of conservation of energy and momentum
across the blast wave (Blandford and McKee 1976).
Although these expressions are
valid only when the blast wave is in the relativistic regime, they are
adequate for our calculation, since X-ray flares usually happen early
before the blast wave enters the trans-relativistic regime.  The amount
of swept up mass in shell of surface area  $4 \pi R^2$ is described by
equation (\ref{density})\footnote{The treatment here is based on the
assumption of an isotropic ejecta. The treatment is valid before the
``jet break'' time, which is usually the case for X-ray flare
observations. After the jet break time, the dynamics may be altered
by sideways expansion of the ejecta. However, numerical simulations
suggest that such an expansion is not important (Kumar \& Granot 2003;
Cannizzo et al. 2004; Zhang \& MacFadyen 2009). For simplicity, we
take the isotropic assumption throughout the blast wave evolution.}.
The solution of this system is found to be
\begin{equation}
\gamma = \sqrt{\f{16 \pi^2 \rho R^6 + 24 \pi \rho (M_0 \gamma_0)R^3 +
9 (M_0 \gamma_0)^2}{16 \pi^2 \rho R^6 + 24 \pi \rho (M_0 \gamma_0)R^3 + 9 M_0^2}}.
\label{bwsolution}
\end{equation}

The leading shell initially expands freely until the momentum of the
swept up matter is about equal to the initial mass of the shell,
$M_0$.  The radius of deceleration, $R_d$, is found from $\gamma_0
\f{4}{3} \pi R^3_d \rho \approx M_0 $, where the subscript ``0''
represents initial values (Rees and M{\'e}sz{\'a}ros, 1992).
For typical values of $\gamma_0 \sim 100$, $M_0 \sim 10^{28}$g and
$\rho \sim 2 \times 10^{-24} {\rm {g}~{cm^{-3}}}$, one has
$R_d \sim 2 \times 10^{16}$cm.

The relativistic blast wave decays as $\gamma\propto R^{-3/2}$,
until reaching the Sedov radius, at
which the rest mass of the ambient medium becomes as large as the
rest mass of the blast wave, i.e. $R_{Sedov}=(3 M_0/ 4\pi
\rho)^{1/3}$.  After this point, the expansion enters the non-relativistic
regime with dimensionless velocity $\beta \propto R^{-3/5}$ (Sedov 1969).

\begin{figure}
\plotone{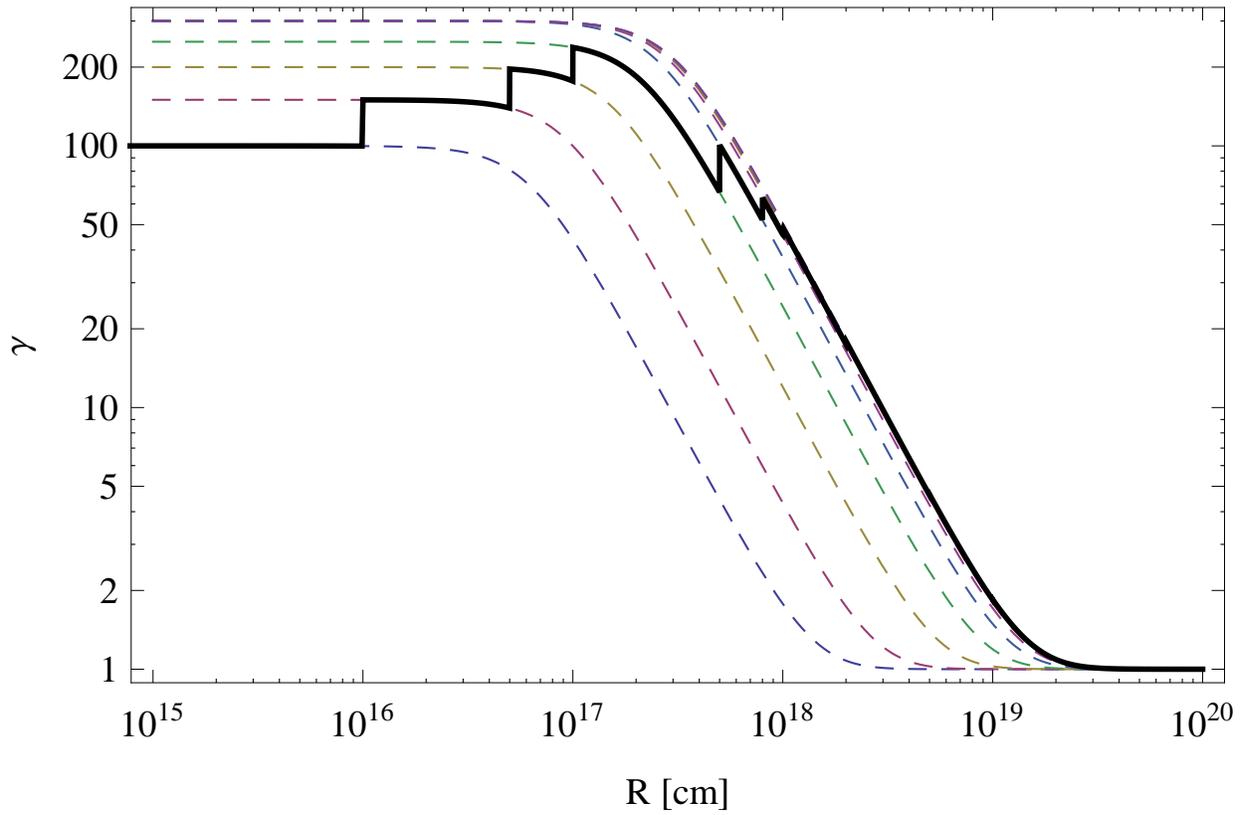}
  \caption{The blast wave shows several glitches as more and more shells pile onto it.
  Successive solutions are shown as dashed lines and the blast wave is shown as a solid line.
  Jumps in the solution become less important as the added energy from a single collision becomes
  less of a fraction of the total energy.}
\label{chopbw}
\end{figure}

In our model, trailing shells collide amongst themselves and
later land onto the
blast wave as the it slows down, altering
the blast wave dynamics. The dynamics of such a collision is
very complicated, invoking three shocks and several
distinct dynamical stages (Zhang \& M\'esz\'aros 2002).
For the purpose of this study (tracking the location of
the blast wave), we adopt the following simple treatment:
If a trailing fast shell with mass $m_{f}$ and Lorentz factor
$\gamma_{f}$ collide on to the blast wave with Lorentz factor
of $\gamma$ before the blast wave has begun decelerating, the collision
is treated as an internal shock and the merged
Lorentz factor are calculated as prescribed by Eq.(\ref{momentum}).
If the collision of a fast shell onto the blast wave occurs after
the blast wave has begun to decelerate, we first assume that there
is no blast wave, and calculate the post-collision product the
same way as the internal shock calculations, record the new
effective initial mass and Lorentz factor, and re-solve the
blast wave for the new parameters. We then jump the blast wave
solution from the old (low) $\gamma$ value to the new (high)
$\gamma$-value at the same $R$.
The code
then tracks this new solution until next collision happens.
With each new collision onto the blast wave, we then calculate
the new solution by changing the effective mass and initial
$\gamma$ factor. By doing so, the blast wave
evolution shows several ``glitches'' with decreasing amplitude,
since the ratio between the trailing shell energy and the
blast wave energy drops with time as the energy of the blast wave
grows (see Figure \ref{chopbw}).

The location $R$ and Lorentz factor $\gamma$ of the blast wave are
traced at any instant $t$. This information is used to screen out
the collisions that are not ``internal''. Without such a blast wave
screening, shells injected from the central engine can collide
at any location at any time. If the relative Lorentz factor of
two shells is small, they can in principle collide at a much larger
radius than the blast wave radius. Should these collisions have happened, both
shells have already entered the ``spreading'' regime, so that the width
of the shells can become very large. This would introduce some
``fat'' X-ray flares which are not observed. The blast wave essentially restricts
such collisions, ensuring that they would never happen. Instead,
both shells collide onto the blast wave and boost up the blast wave
energy. The inclusion of the blast wave dynamics is therefore
essential, screening out many ``fat'' flares. This also
explains the lack of very ``fat'' flares in the X-ray afterglow
data.

In order to reproduce the data, we
adopt a typical afterglow template
according to the data (Nousek et al. 2006; O'Brien et al. 2006; Liang et
al. 2007) to denote the underlying X-ray afterglow. This is a broken
power law with decay indices -1/2 and -1.25, with a break time $t_b
=6 \times 10^3$s and a break flux ${\rm Flux}(t_b)=1.3 \times 10^{-11} ~{\rm
erg~s^{-1}~{cm^{-2}}}$.  The origin of the X-ray afterglow, especially
the shallow decay phase, still remains a mystery. A number of very
different physical mechanisms have been proposed to explain its origin
(see Zhang 2007 for a review). Although the refreshed shock model
is mostly discussed (Zhang et al. 2006; Nousek et al. 2006;
Granot \& Kumar 2006), some optical afterglows do not
show a similar temporal break around the X-ray break time (Panaitescu
et al. 2006; Liang et al. 2007) suggesting that this model cannot
interpret all the X-ray afterglow data. Other ideas/models
include a central engine powered afterglow (Ghisellini et al. 2007;
Kumar et al. 2008), a long-lasting reverse-shock-dominated afterglow
(Genet et al. 2007; Uhm \& Beloborodov 2007), two-component
external shock (Racusin et al. 2008; De Pasquale et al. 2009), dust
scattering (Shao \& Dai 2007), up-scattering of blast wave photons
by a trailing lepton-rich ejecta (Panaitescu 2008) and an emission
component prior to the GRB trigger (Yamazaki 2009; Liang et al. 2009).
Since none of the scenarios have been robustly proven, we do not
demand our model to interpret the power law X-ray afterglow segments
self-consistently.

On the other hand, the expected external shock X-ray afterglow flux
can be calculated from our model. Applying the standard afterglow
model (Sari et al. 1998), we can calculate the X-ray lightcurve
for a particular simulation.
Calculating the flux for either the fast cooling case
\begin{equation}
F_{\nu}  = \left\{
\begin{array}{l l}
  \left( \nu/\nu_c \right)^{1/3} F_{\nu , max} & \quad \nu_c \geq \nu \\
  \left( \nu/\nu_c \right)^{-1/2} F_{\nu , max} & \quad \nu_m \geq \nu \geq \nu_c \\
   \left( \nu_m/\nu_c \right)^{-1/2} \left( \nu/\nu_m \right)^{-p/2} F_{\nu , max} & \quad \nu \geq \nu_m \\ \end{array} \right. \
\label{fastcooling}
\end{equation}
or slow cooling case
\begin{equation}
F_{\nu}  = \left\{
\begin{array}{l l}
  \left( \nu/\nu_m \right)^{1/3} F_{\nu , max} & \quad \nu_m \geq \nu \\
  \left( \nu/\nu_m \right)^{-(p-1)/2} F_{\nu , max} & \quad \nu_c \geq \nu \geq \nu_m \\
   \left( \nu_c/\nu_m \right)^{-(p-1)/2} \left( \nu/\nu_c \right)^{-p/2} F_{\nu , max} & \quad \nu \geq \nu_c \\ \end{array} \right. \
\label{fastcooling}
\end{equation}
using the minimum Lorentz factor of the electrons
$\gamma_m = \epsilon_e \left ( \f{p-2}{p-1} \right ) \f{m_p}{m_e} \gamma$,
the comoving magnetic field strength $B = (32 \pi m_p \epsilon_B n)^{1/2} \gamma c$,
and the cooling Lorentz factor of electrons
$\gamma_c = \f{6 \pi m_e c}{\sigma_T B^2 \gamma t} \simeq \f{6 \pi m_e
  \gamma c^2}{\sigma_T B^2 R}$,one can calculate the critical synchrotron frequencies $\nu_m$ and $\nu_c$ using
$\nu (\gamma_e) = \gamma \gamma_e^2 \f{q_e B}{2 \pi m_e c}$.
Standard values are taken for $p = 2.4$, $\epsilon_e = 0.1$,
$\epsilon_B = 0.01$, $n = 1$ (Panaitescu \& Kumar 2002)
and $m_p$, $m_e$, $\sigma_T$ and $q_e$
are the proton mass, electron mass, fundamental charge and Thompson
cross-section, respectively.   Combining this with
\begin{equation}
F_{\nu,max} = \f{m_e c^2 \sigma_T}{3 q_e} \gamma n B  R^3 \f{ \nu}{4 \pi D^2},
\label{Fmax}
\end{equation}
where $D$ is the distance from the observer, here taken to be $10^{28}$
cm, $\nu$ is the frequency in the X-ray band, taken to be $10^{18}$
Hz, one can calculate the X-ray flux density, $F_{\nu}$ as a function
of radius taking care to switch between fast and slow cooling cases where
appropriate (Sari, Piran \& Narayan 1998).

\begin{figure}
\plotone{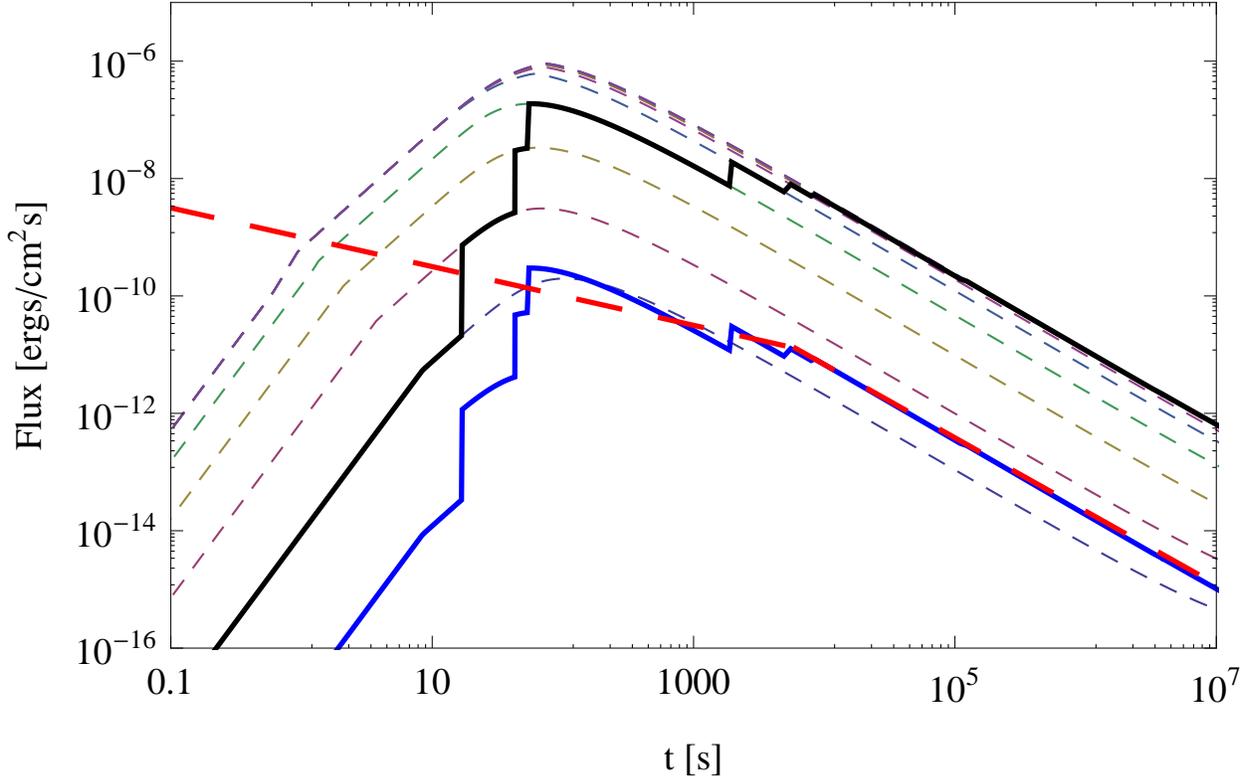}
  \caption{The flux lightcurves are shown for the same blast wave used in
Fig.\ref{chopbw}, calculated based on the external shock model. The
calculation of $\epsilon_e=0.1$ is displayed in detail: the light dashed
lines are the successive blast wave solutions, and the upper solid line is the
corresponding blast wave lightcurve, which jumps between solutions.
These glitches occur as shells pile onto the blast wave and decrease in magnitude as
the energy of added shells becomes less significant. The lower solid
line is the lightcurve for $\epsilon_e = 10^{-3}$, which matches the
template adopted in the rest of the calculations (thick dashed line).}

\label{glitchplot}
\end{figure}

Shells colliding onto the blast wave will produce ``glitches'' in the
regular afterglow flux decay.  Successive solutions, shown as dotted
lines in Fig.\ref{glitchplot}, depend upon the energy contained in the
blast wave and therefore on both the Lorentz factor, $\gamma_0$, and
$M_0$, the effective mass of the blast wave.

We use the calculated $\gamma$-evolution to calculate the evolution
of the X-ray flux. In order to plot the flux as a function of time
instead of radius, points are plotted as abscissa $t = R/(2 c
\gamma^2)$ and ordinate flux where both flux and gamma are calculated
by matching radius.  Since the blast wave is accelerating in the rising
part of Fig.\ref{glitchplot}, here points appear to move back in
time as the next solution is taken.  In this part of the plot,
blast wave solutions have simply been connected vertically.  In the
falling part of Fig.\ref{glitchplot}, solutions appear to move
forward in time as collisions occur.  In general, all these
abrupt jumps are artificial. In reality, one needs to consider
the effect of equal arrival times, which smear out all the abrupt
jumps so that the lightcurve would appear smoothed without
noticeable individual glitches.

Also shown in Fig.\ref{glitchplot} are another blast wave solution
for $\epsilon_e = 10^{-3}$, and the afterglow template adopted
in other calculations throughout the paper. As evident from the
figure, for the standard value $\epsilon_e = 0.1$ as derived from
broad band afterglow modeling (e.g. Wijers \& Galama 1999;
Panaitescu \& Kumar 2002; Yost et al. 2003), the X-ray flux predicted
by the external shock model out-shines the template flux level
(which is based on observations) by about three orders of magnitude.

\begin{figure}
\plotone{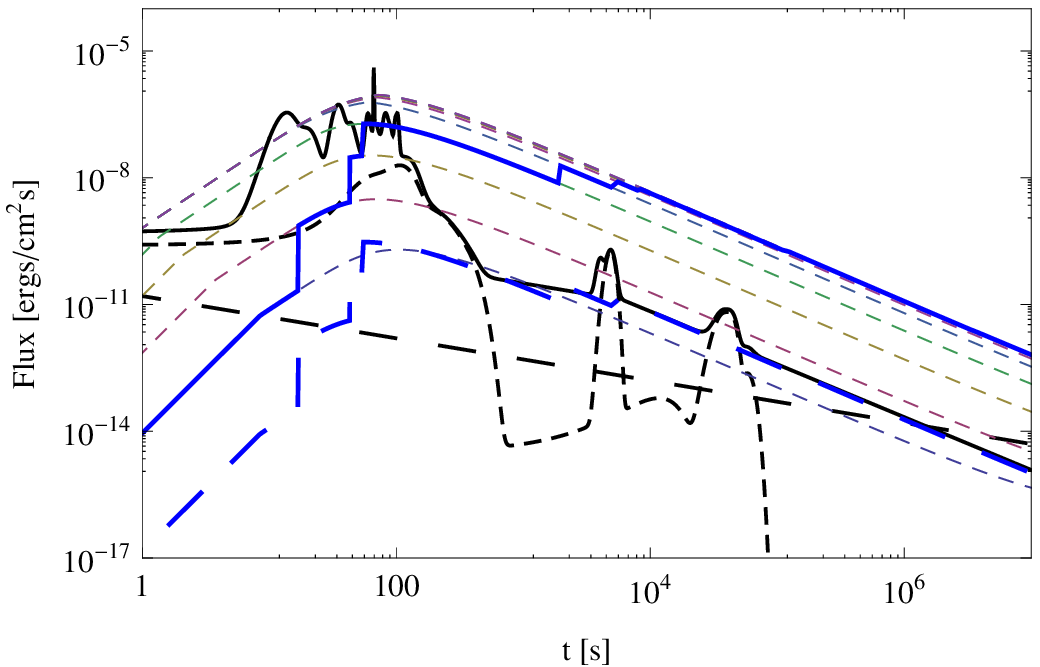}
  \caption{The calculated X-ray lightcurve including prompt emission,
X-ray flares and a broken power law template (solid black line), as
compared with the lightcurves calculated from the external shock
blast wave model (two colored lines: $\epsilon_e = 0.1$ for the
upper solid one and $\epsilon_e = 10^{-3}$ for the lower dashed one). }
\label{glitchfull}
\end{figure}

This is also emphasized in Fig.\ref{glitchfull} which compares the
simulated lightcurve (including the contributions from the
prompt emission and X-ray flares superimposed on the template)
and the calculated external shock lightcurve.
It is evident that no steep decay
phase and X-ray flares are observable if $\epsilon_e = 0.1$ is
adopted to calculate the afterglow level. This issue is carried over
from the low efficiency problem of the internal shock model.
If the observations are to be reproduced, there are two possibilities.
The first is that the blast wave radiation efficiency is much lower.
We test this possibility by lowering $\epsilon_e$, and found that
the predicted afterglow level can be roughly reproduced if
$\epsilon_e$ is as low as $10^{-3}$ (Figs.\ref{glitchplot} and
\ref{glitchfull}).
The pile-up effect of shells onto the blast wave naturally produces
a shallow decay phase (after smoothing the abrupt jump features),
which is an attractive feature of
this internal-external-shock model. However, the anomalously small
$\epsilon_e$ is inconsistent with the values derived from the previous
broadband afterglow modeling, suggesting that this is likely not the
correct approach to solve the problem. The second possibility is that
the radiative efficiency of the prompt gamma-ray emission
is very high (e.g. Zhang et al. 2007).  This requires a more efficient
mechanism to generate the prompt emission. The data of the recent GRB
080916C (Abdo et al. 2009)
suggests that the outflow is very likely Poynting flux dominated
(Zhang \& Pe'er 2009). Within such a picture, Zhang \& Yan (2009)
proposed an Internal Collision-induced MAgnetic Reconnection and
Turbulence (ICMART) model, which retains the merits of the internal
shock model but significantly increases the prompt emission efficiency.

The time history analysis of our shell model is also applicable to the
ICMART model. The prompt emission and X-ray flare features can be
retained, but the radiation efficiency is increased. Within such a
scenario, the external shock level can be lowered to satisfy the
observational constraint.

One can estimate the amplitude of ``glitches'' in the X-ray light curves.
In general, collisions onto the blast wave are not energetic enough to
produce a prominent signature on the afterglow light curve, unless
the injection energy is comparable to that already in the blast wave
(Zhang \& M{\'e}sz{\'a}ros, 2002). In any case, small glitches from
arriving shells, although not individually seen, could effectively
bump up the normal decay phase, making it appear shallower.
Postcollisions of shells sorted by decreasing Lorentz factor by
internal collisions could produce an observable signature in early
afterglow lightcurves and may be seen as temporal variability or a
deviation from power-law decay  (Kumar \& Piran, 2000). The amplitude
of glitches produced by postcollisions can be calculated knowing the
initial mass and Lorentz factor of both the blast wave and colliding
shell and will essentially represent how far in the vertical direction
successive solutions are separated.  For the first few postcollisions,
the amplitudes of the glitches are rather large as the added shells
have both Lorentz factor and mass comparable to that of the blast wave
itself.  In this first stage, there is no simple approximation for
calculating the jump in solution.  The percent increases
($(F_f-F_i)/F_i$) are 3000\%, 1000\%, 500\% and 150\% for the presented
simulation. During the deceleration phase, the X-ray flux density $F_\nu(X) \propto E^{(p+2)/4}$
for $\nu_x > {\rm min}
(\nu_m, \nu_c)$ and $F_\nu(X) \propto E^{(p+3)/4}$ for $\nu_m < \nu_x
< \nu_c$, where $E$ is the total energy in the blast wave. The glitch
amplitude is simply determined by the increase of the blast wave energy
during each collision. During the collision, there is a reverse shock
propagating to the trailing shell. However, this reverse shock usually
does not contribute significantly into the X-ray band, since its
density is higher, and hence, the typical electron Lorentz factor is
much lower. Its dominant output is in the optical band (M\'esz\'aros
\& Rees 1997; Sari \& Piran 1999; Zhang \& M\'esz\'aros 2002).

It has been argued that for GRBs with large peaks followed by deep
troughs could be used to put an upper limit on the value of $\gamma_0$
(Zou \& Piran, 2009).  The external shock component will be
superimposed on the internal shock prompt pulses and if this external
shock does not rise above the level of the trough and/or threshold
sensitivity of the detector at that point, this could be used to put
an upper limit on the Lorentz factor of this first blast wave shell by
knowing how high successive blast wave solutions will rise in flux.  We
find that this method, in principle, is a feasible method of putting a
upper limit on $\gamma_0$ with a few caveats in mind.  Figure
\ref{glitchfull} shows the blast wave solutions and blast wave (external
shock component) plotted with the internal shock component lightcurve
for one simulation.  In general, the first shell of the external shock
component will not be energetic enough to rise to the level of the
prompt emission.  As more and more shells pile onto the blast wave, the
flux level rises and so looking for a solution that could be
extrapolated backwards to find the initial $\gamma_0$ is problematic
since the external shock component, especially early on, will not be a
smooth function of time. Another complication is that the flux level
of the blast wave emission sensitively depends on the unknown
$\epsilon_e$ parameter, making the derived upper limit of $\gamma_0$
subject to large uncertainty.

\section{Multiple Shell Simulations}
\subsection{Single Injection Episode}
\begin{figure}
\plottwo{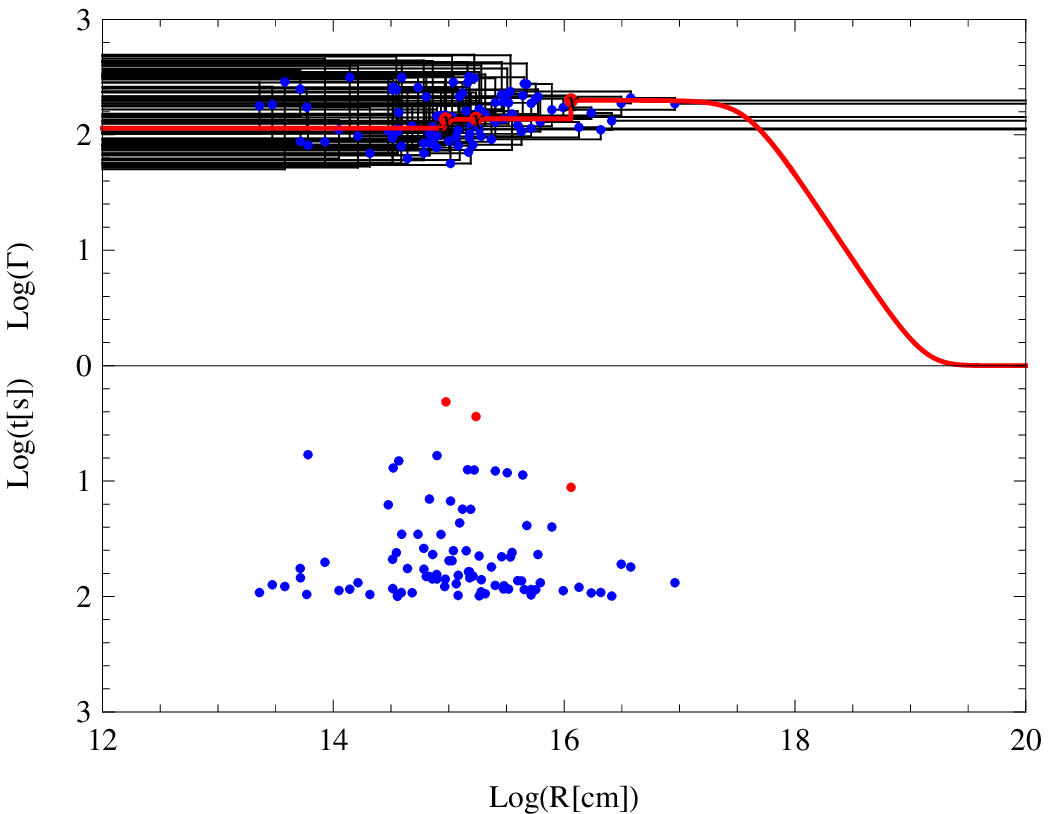}{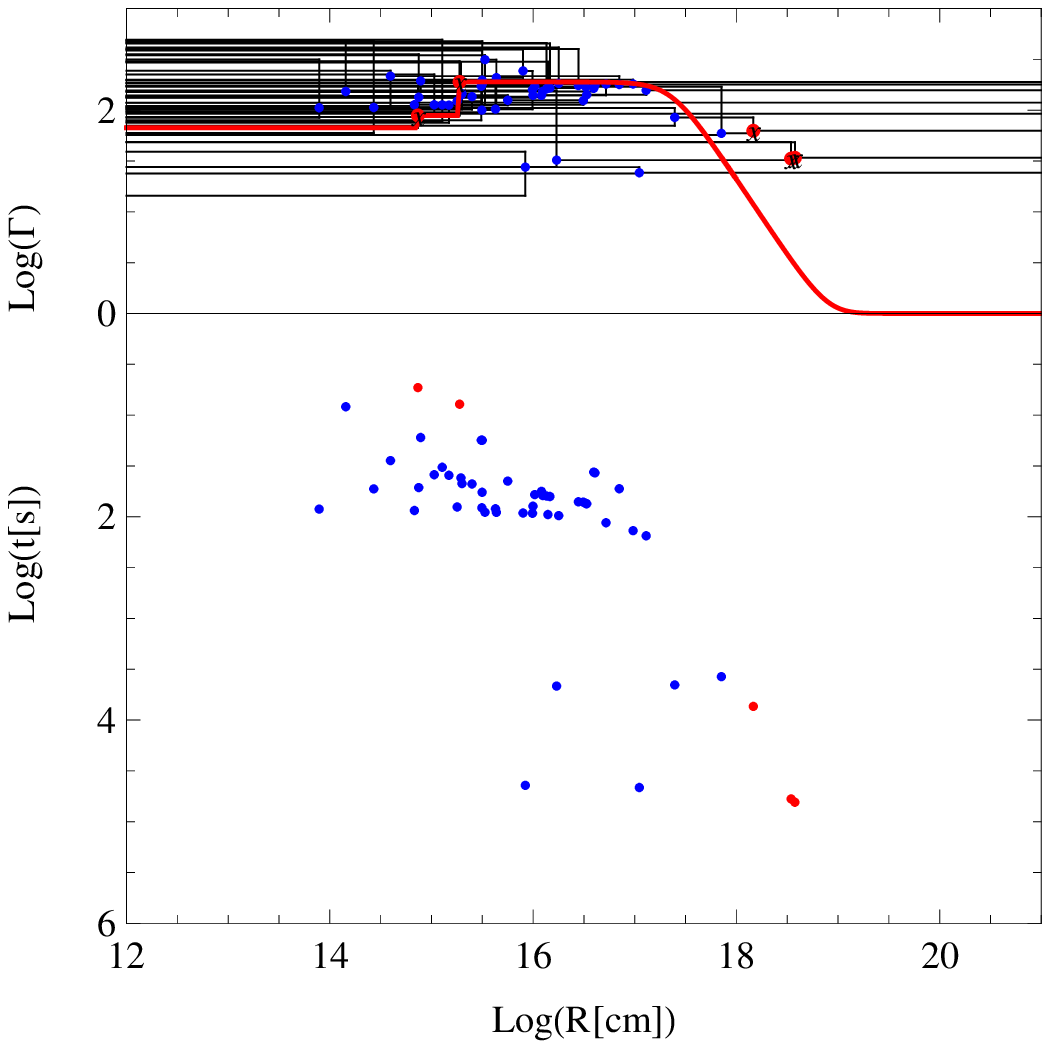}
  \caption{Tree plots showing simulated collisions as both Lorentz
factor versus radius (top panes)
and time of collisions versus radius of collisions (bottom
panes).  Lorentz factors are read up the y-axis starting at zero,
time of collision is read down the y-axis, also starting at zero.
The top pane shows shells as black lines, collisions are marked with
dots.  The time of collision can be found by dropping
down vertically to the bottom pane of the graph, matching radius.
The blast wave is shown in the
top panels as a thick red line.  Allowed collisions are blue
dots, excluded collisions are red dots. For the left pane, 100 shells are ejected between 0 and 100 seconds
with Lorentz factors between $50-500$. For the right pane, 50 A shells are ejected between $0<t_{ej}< 100$
    seconds, 5 B shells $3000<t_{ej}< 5000$ and 5 C shells
    $30000<t_{ej}< 50000$ seconds.}
\label{treeplot100shells}
\end{figure}

\begin{figure}
\plottwo{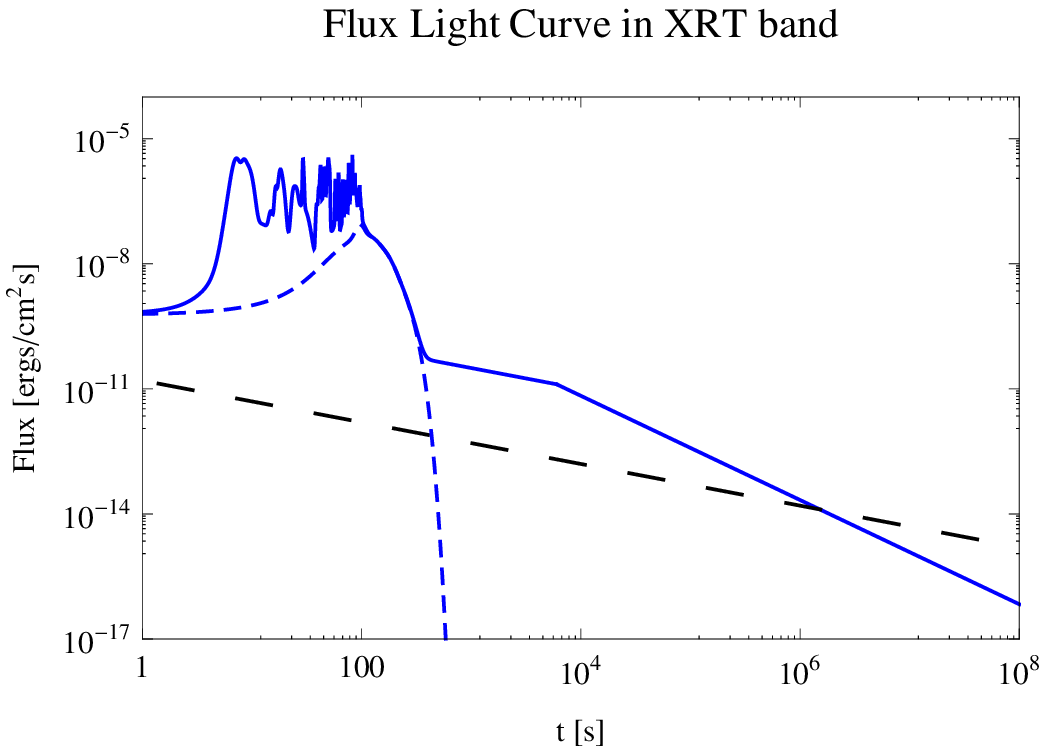}{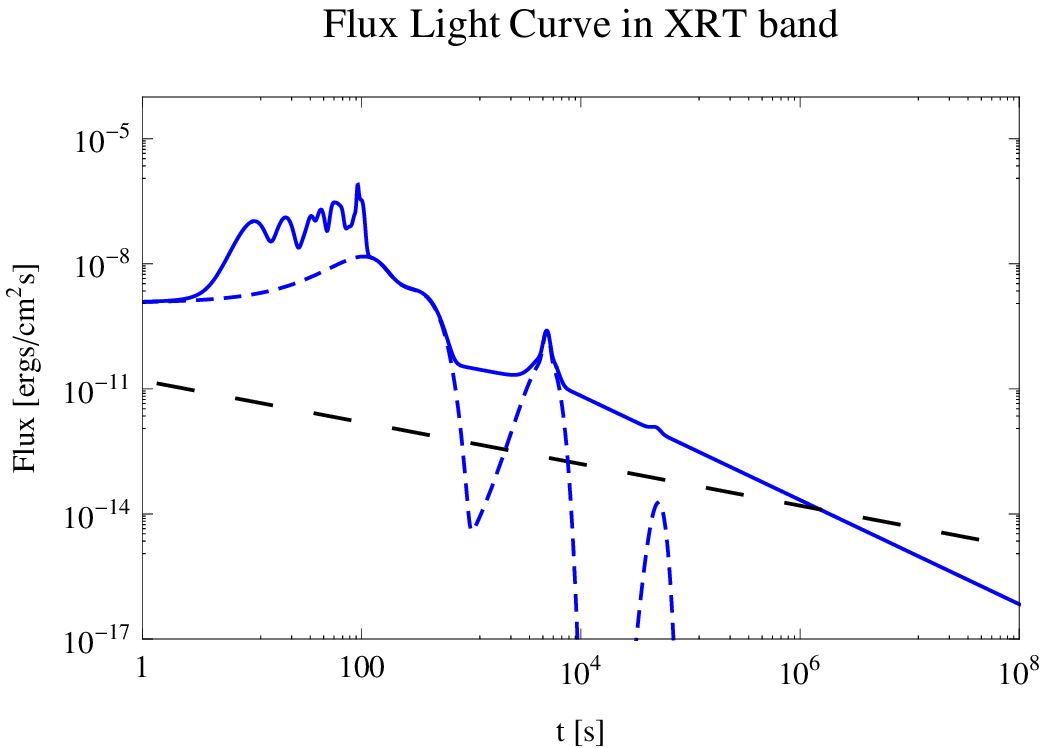}
  \caption{A simulation with many shells in a single short episode
    will produce prompt emission only and no late collisions which
    could be responsible for X-ray flares.  Left panel is a single
    ejection episode N = 100, $0<t_{ej}<100$ seconds.  Right panel a
    simulation of N = 50 A shells, $0<t_{ej}<100$ seconds, 5 B shells
    $3 \times 10^3<t_{ej}<5 \times 10^3$ seconds and 5 C shells $3
    \times 10^4 <t_{ej}<5 \times 10^4$ seconds.  Dashed lines show the
    position of flares underneath the superimposed afterglow.}
  \label{prompt}
\end{figure}

We first model collisions of a group a shells that are injected in
a single emission episode. This is relevant to GRBs that have prompt
emission without
distinct gaps between pulses. Allowing the central engine to eject
multiple randomized shells which go on to collide, we track the
information of individual shells. An example is shown in
Figs.\ref{treeplot100shells}a and \ref{prompt}a. In this simulation,
100 shells are ejected where shell initial thickness, Lorentz factor and
mass are chosen from random distributions in log space:  Lorentz
factors:  $50  < \gamma < 500$, mass: $10^{29} < m < 10^{31}$,
initial thickness:  $10^{10}< \Delta_0 < 2 \times 10^{10}$, all in
cgs units. Ejection times from the central engine are chosen from a
linear random distribution  $0 < t_{ej} < 100$ seconds in the
rest frame of the GRB central engine.
When two shells collide, we let them merge, drop one shell, and adopt
the merged shell parameters ($m$, $\gamma$, $\Delta$) as the new
values of the remaining shell. In order to keep track of future
collisions, we also need to re-set the ``effective'' ejection time
of this new shell, which is taken as $t_{ej,m}=R_{col}/c\beta_m$,
where $\beta_m=(1-1/\gamma_m^2)^{1/2}$. The code then runs again
with one shell reduced. The same procedure is applied when each
collision happens, so that the code can track all the
collision/merging processes for any arbitrarily designed central
engine activity.

\begin{figure}
\plotone{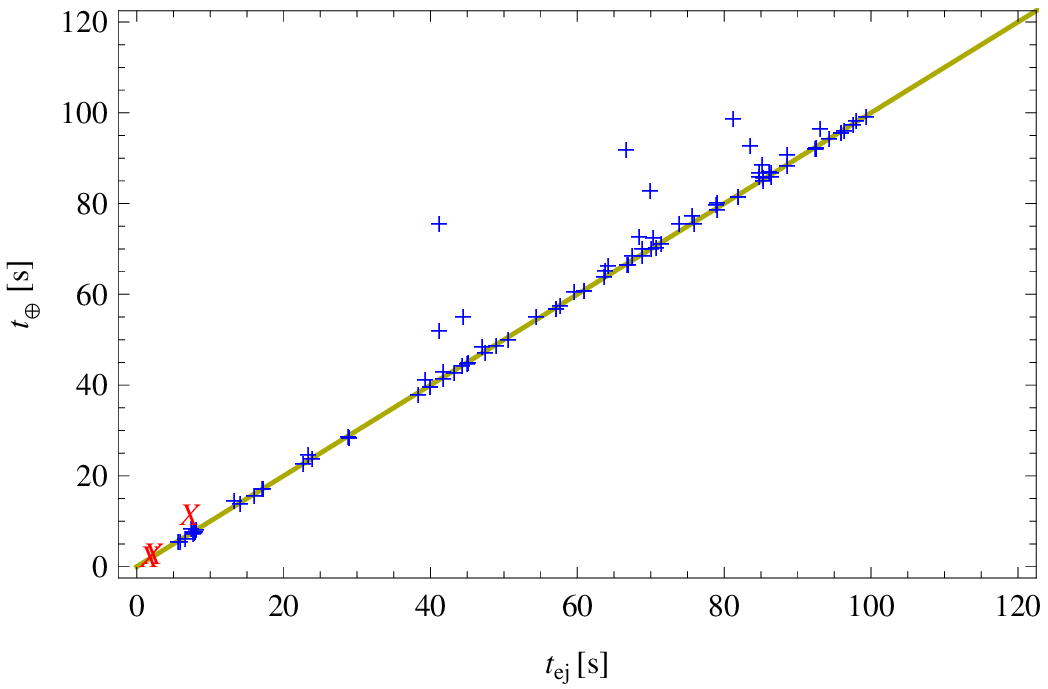}
  \caption{The ejection time of a shell in the GRB central engine rest
    frame ($t_{ej}$) versus the collision time in the observer frame
    $t_{\oplus,col}$.  Allowed collisions are pluses, crosses are
    excluded.  Even for a large number of shells (N = 100, ejected in
    100 seconds), there is little scatter around a line of unity.}
\label{corrplot2}
\end{figure}

Figure \ref{treeplot100shells}a displays the ``tree-plot'' of this
simulation. The upper panel displays how shells with different
Lorentz factors are ejected, collide and merge at various
distances. The time information is not displayed, but the
collision times $t_{col}$ (again in the rest frame of the central
engine) can be read off from the lower panel. This time is
translated into the observed time according to Eq.(\ref{time})
to calculate the lightcurve (Fig.\ref{prompt}a). The evolution
of the blast wave is also marked in the upper panel as a thick red line.
Collisions are ``disallowed'' if the shells collide with the blast wave (i.e.
they would have collided at a radius greater than that of the blast wave,
had the blast wave not existed.). These shells are
included to boost the blast wave energy and become part of the
blast wave after the collision and are therefore no longer traced in
the later simulation. Our results suggest that a single
episode can reproduce prompt emission of some GRBs.
Even for simulations with a large number of shells, the ejection
time in the frame of the central engine ($t_{ej}$) is correlated
to the observed time of the collision ($t_{\oplus,col}$) with little
scatter (Fig. \ref{corrplot2}, see also Kobayashi et al. 1997).
There are indeed collisions with
much larger collision times (red crosses), but they are ``excluded''
by the blast wave constraint. If prompt emission and X-ray flares
are indeed produced by the same mechanism, late flares are unlikely
to be produced by shells that are ejected early. As discussed in
the following, they demand re-activation of the central engine.
Lightcurves are produced by considering the spectral model and
temporal model as discussed in \S2. The XRT band lightcurve of
this particular simulation is shown in Fig.\ref{prompt}a.  The
relevant detector threshold (thick, dotted line) is also added.
(D. N. Burrows, 2008, private communication). It is evident that
without late central engine activity, no X-ray flares can be detected.

\subsection{Multiple Injection Episodes}

\begin{figure}
\plotone{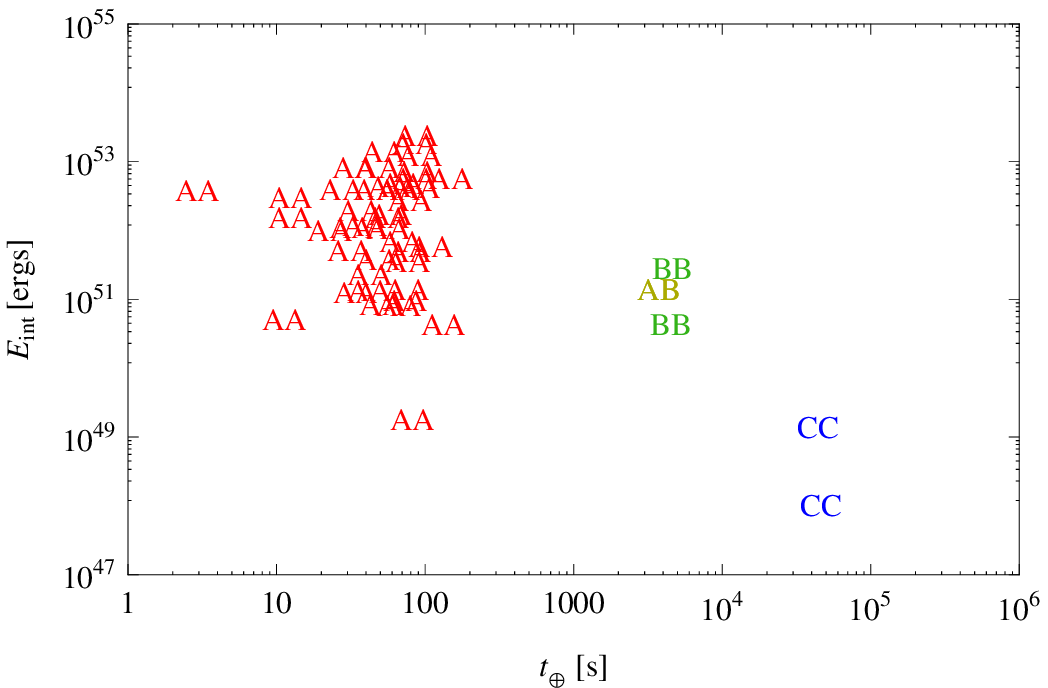}
  \caption{A bolometric light curve for 50 A shells ejected within the
    first 100 seconds, followed by 5 B shells after 3000 seconds and
    5 C shells after 30,000 seconds representing a restarting of the
    central engine. Collision energies are shown by type; ``AA''
    represents collisions between to A shells, ``AB'' between an A
    shell and a B shell etc.}
\label{bololightcurve}
\end{figure}

\begin{figure}
\plottwo{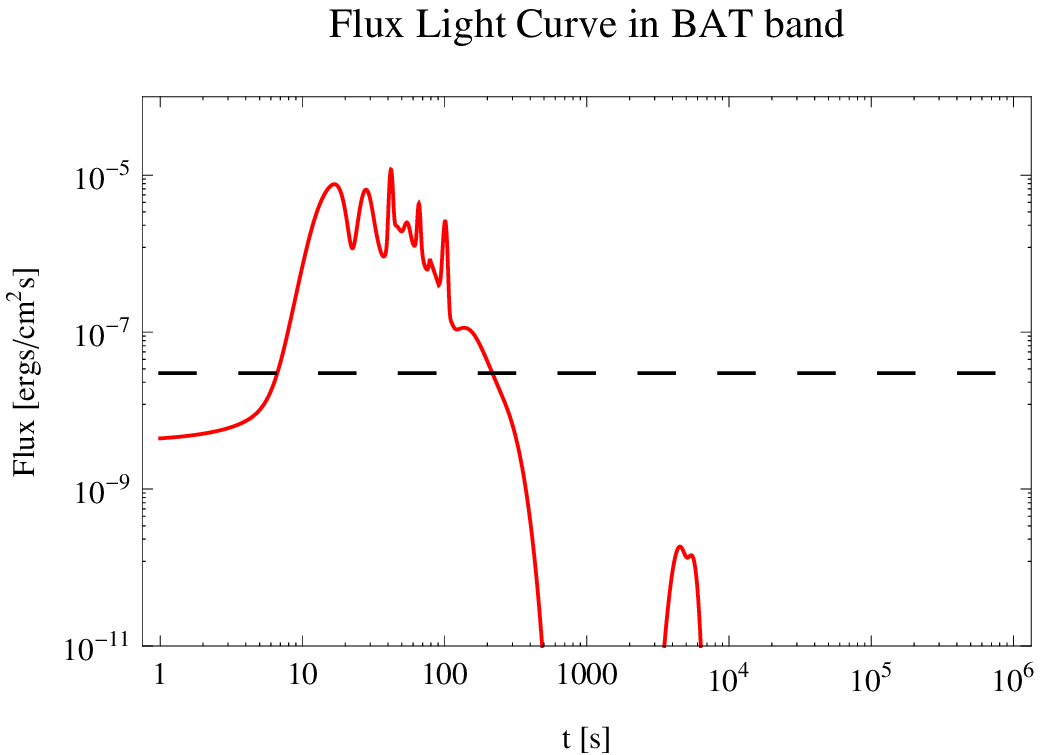}{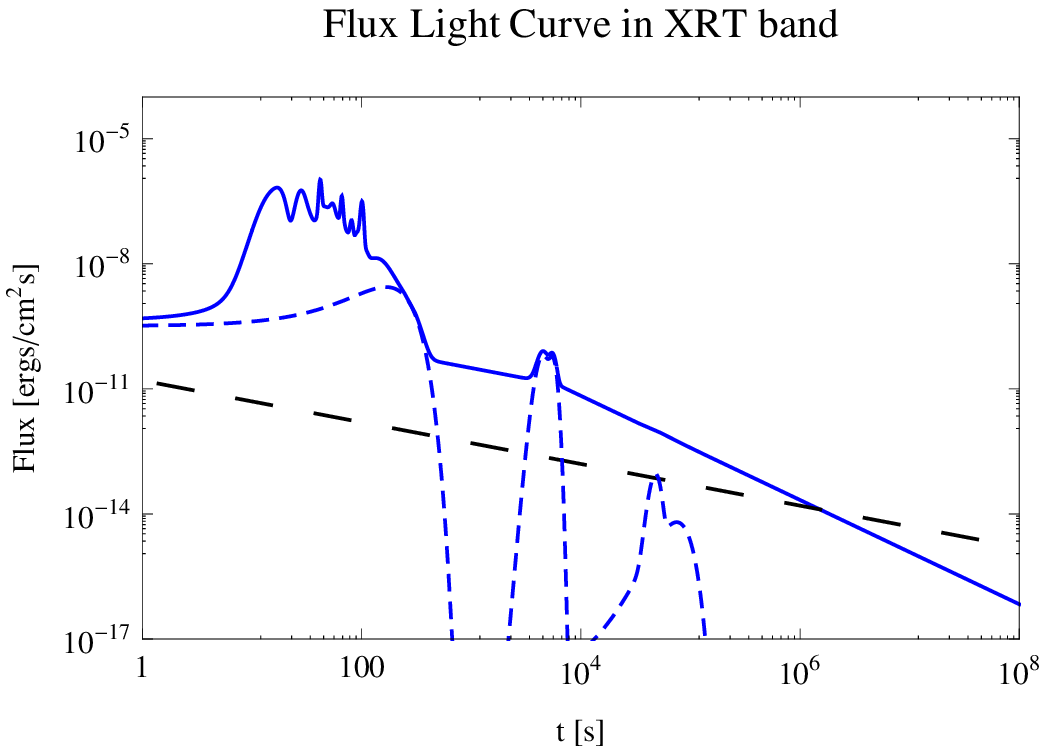}
\plottwo{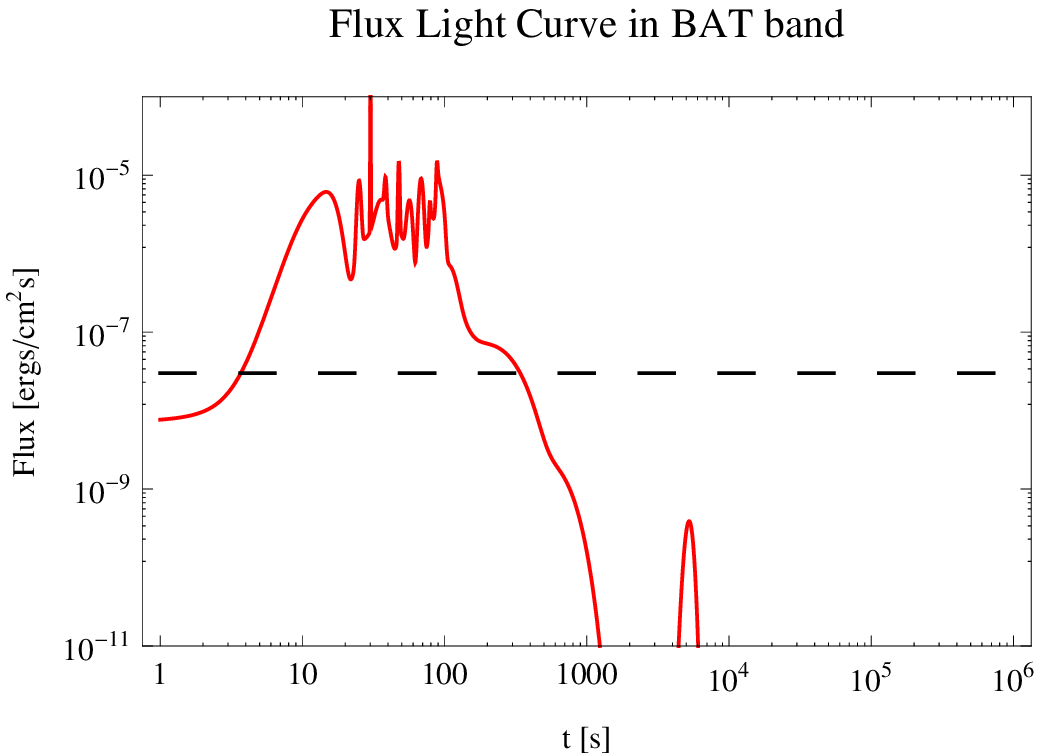}{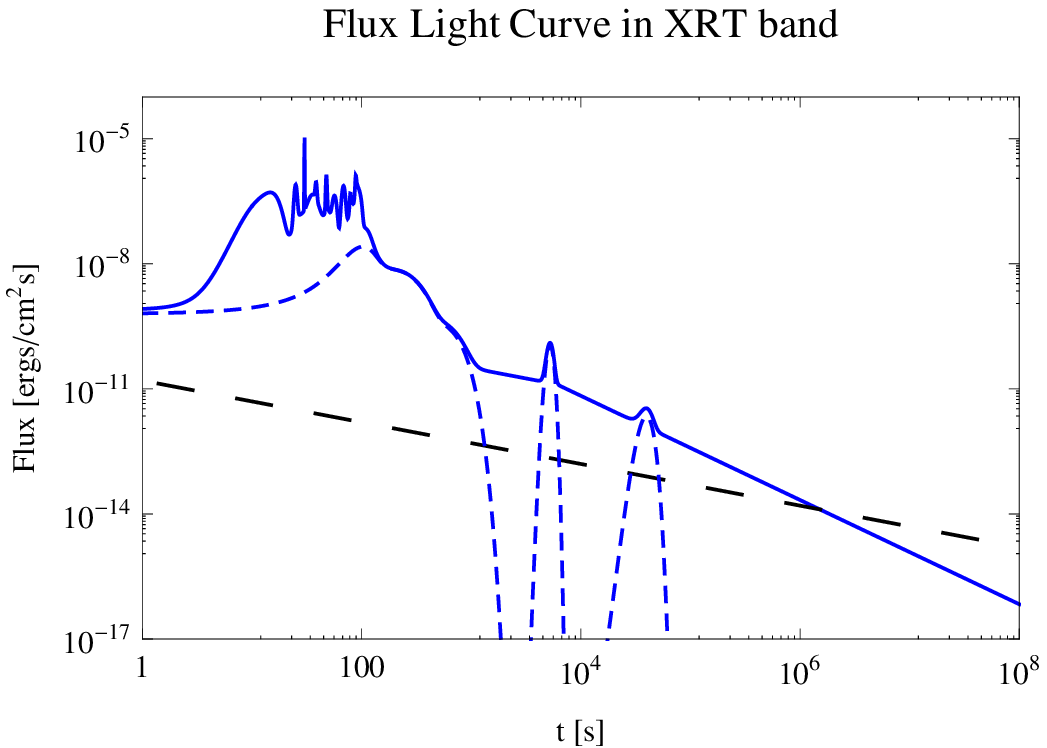}
\plottwo{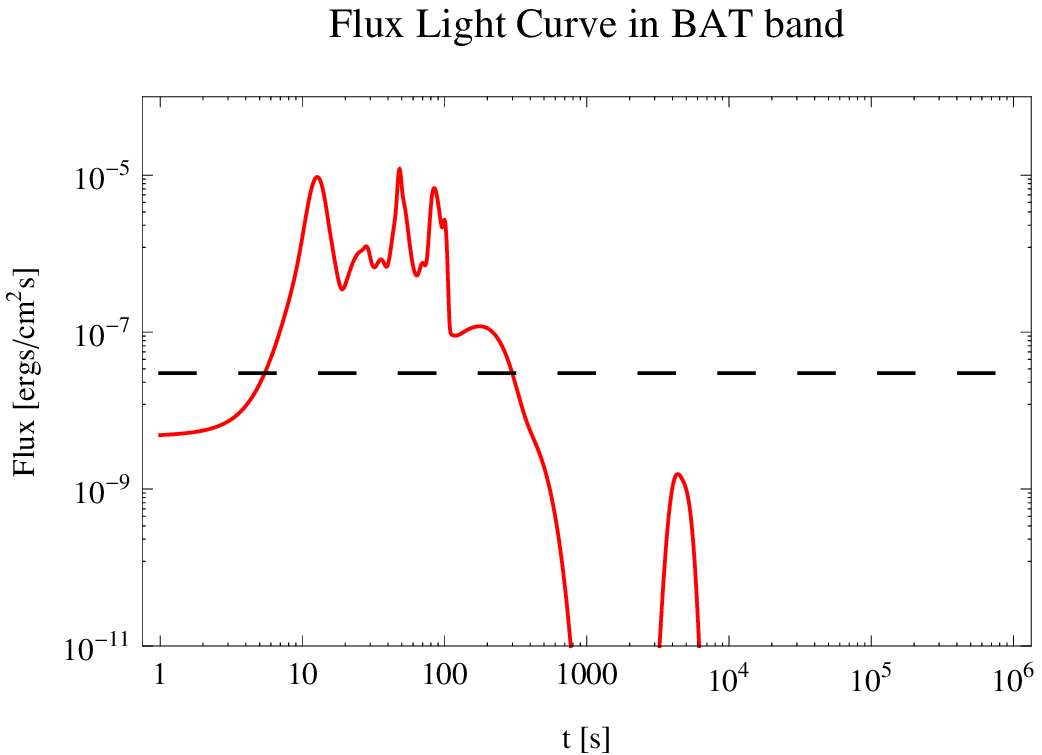}{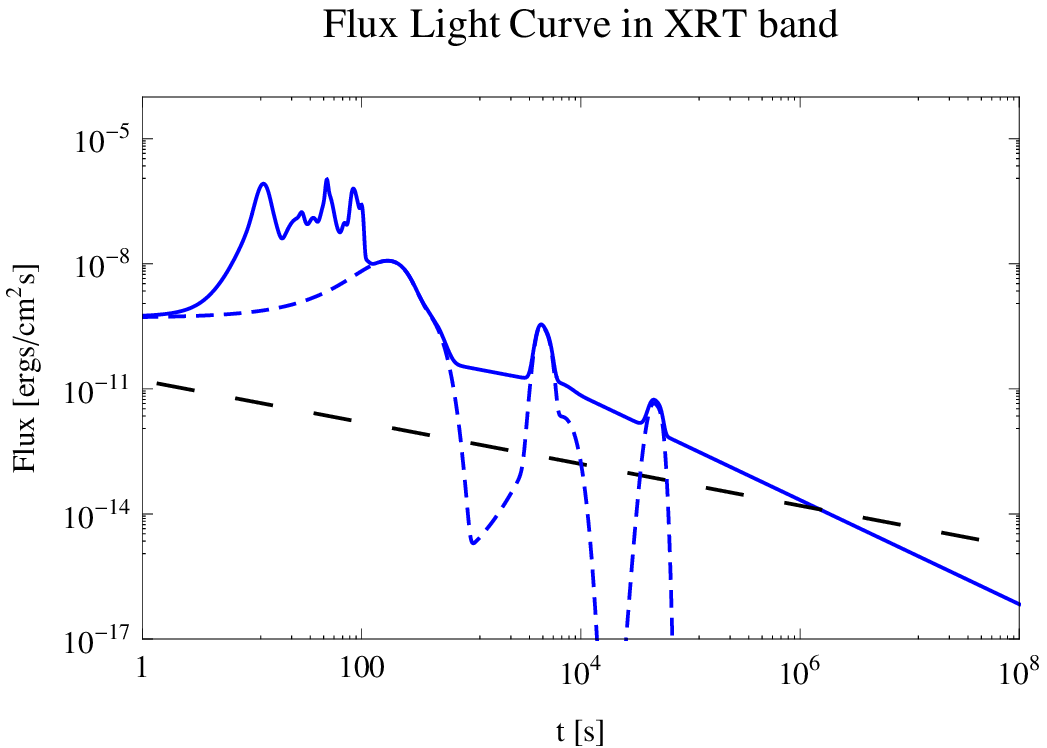}
    \caption{BAT (left) and XRT (right) light curves for three typical
      simulations of 50 A,  5 B shells and 5 C shells of decreasing
      energy. Long dashed lines are the detector thresholds and short
      dashed lines in XRT lightcurves represent the positions of the
      flares underneath the afterglow.}
\label{fluxlightcurves}
\end{figure}

In order to reproduce the observed X-ray flares that occur in distinct
emission episodes from the prompt emission, we are obliged to simulate
multiple ejection episodes from the central engine. The results are
displayed in Figs. \ref{bololightcurve}, \ref{treeplot100shells}b,
\ref{prompt}b and \ref{fluxlightcurves}.  In these simulations, three
groups of shells (``\textit{A}'', ``\textit{B}'' and ``\emph{C}'') are
released.  The \textit{A} group consisting of 50 shells is released in
the first 100 seconds. The majority of these shells merge with each
other producing \textit{AA} type collisions. The energy released in
each collision is shown as ``AA'' in Fig.\ref{bololightcurve}.
Similar to the previous simulation (\S4.1), most \textit{AA} type
collisions occur before 100 seconds.  After 3000 seconds, we release a
set of 5 \textit{B} group shells with energies reduced by $\sim$ 20 times
on average. These shells can have two types of collisions: \textit{BB}
type collisions  or \textit{AB} type collisions (see
Fig.\ref{bololightcurve}).  Both the \textit{AB}
and \textit{BB} type collisions could produce X-ray flares,
as long as they are bright enough to stick out above the power law
afterglow level (the broken power law template).
For late X-ray flares, the energy
of these collisions is of the ``goldilocks type'':  high enough to be
above the background decay, but small enough to peak in the X-ray band
and therefore remain undetectable by BAT.
After 30,000 seconds, a third batch of \textit{C} shells are released
with energy lower than that of the \textit{B} shells by  $\sim$ 20 times
again. This can in principle produce \textit{CC,AC} or \textit{BC} collisions,
but for this particular simulation, \textit{AC} and \textit{BC} collisions are lacking
(Fig.\ref{bololightcurve}). These cross collisions are disfavored
since their collision radii tend to be large (due to the large
ejection gap between the shells), so that the leading shells likely
have collided onto the blast wave before the trailing
shell catches up. Progressively less energy in \textit{A},
\textit{B} and \textit{C} shells allows progressively
degrading energetics of the X-ray flares, as
is commonly observed. This is also generally consistent with various
central engine models, where the accretion or magnetic power of the
engine tends to die off with time.

\begin{figure}
\plottwo{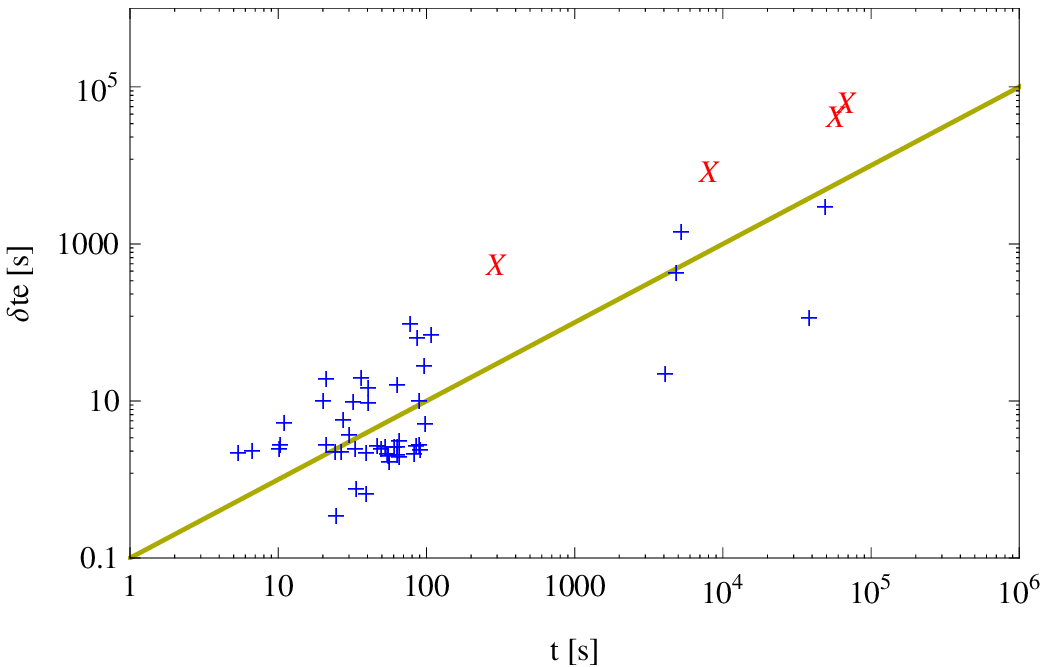}{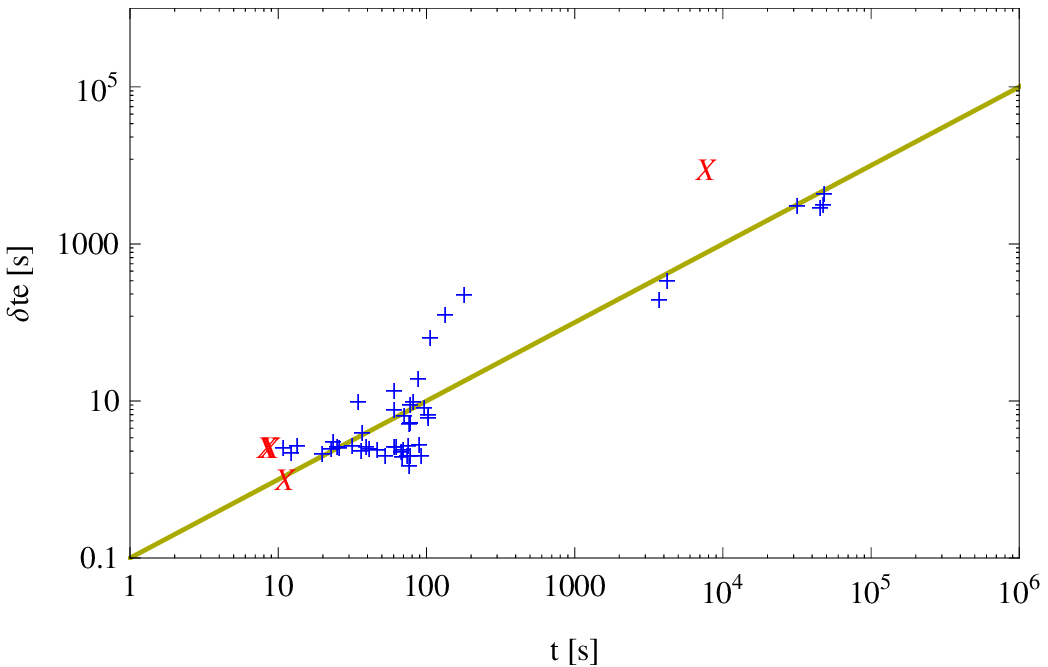}
  \caption{How fat are the flares compared to the time of their
    occurrence?  By the time most shells have spread, creating a broad
    flare, these shells have already collided with the blast wave or
    are not energetic enough to be seen. Observed flares are, on
    average, 1/10 of their occurrence times in width, this line is
    shown. Excluded collisions are shown as crosses, allowed
    collisions are pluses.  The left panel is for shells of the same width, the
    right for shells which are wider for later ejection episodes.}
\label{widthinfo}
\end{figure}

\begin{figure}
\plottwo{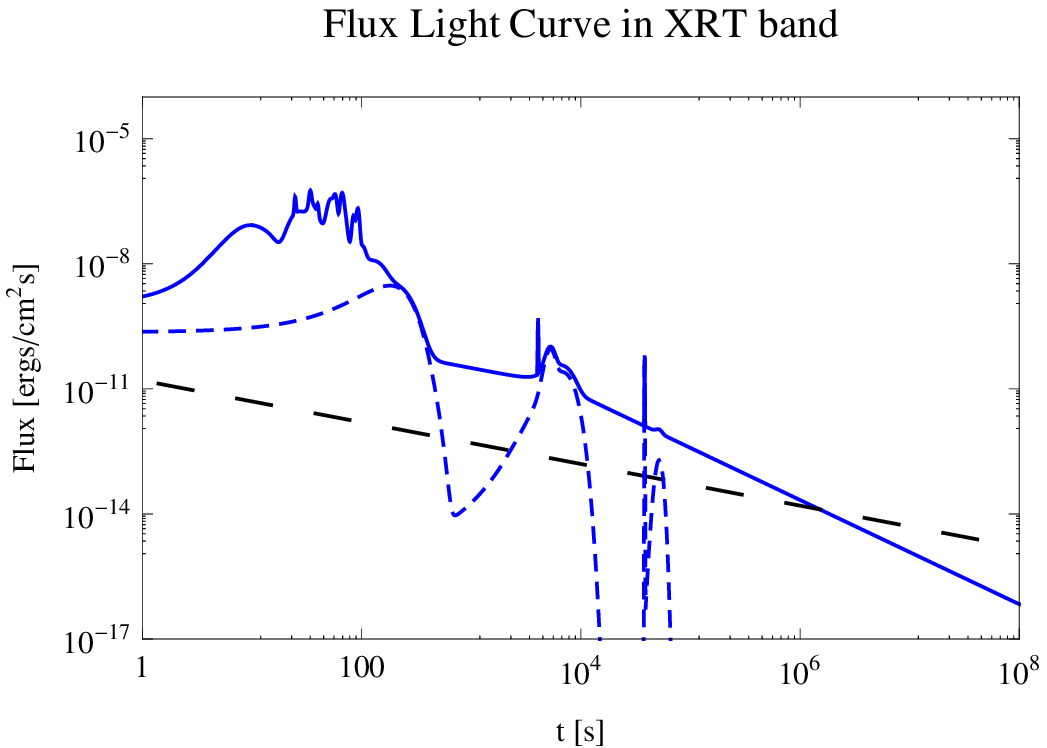}{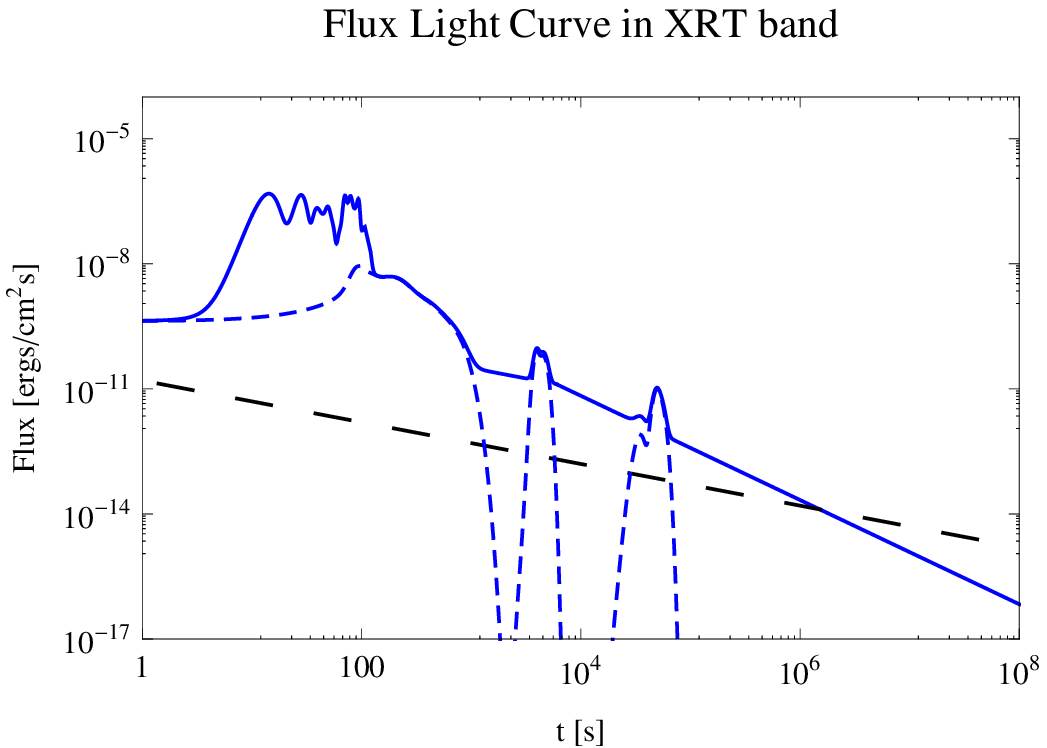}
  \caption{Simulations with uniform shell widths (left) produce flares
    which are more narrow on average than observed flares. In order to
    reproduce observed wide flares, shells must be ejected with widths
    scaling as their ejection time (right panel).  Notations are the
    same as for Fig. \ref{fluxlightcurves}}
\label{latefatshells}
\end{figure}

Another interesting topic is to investigate how the observed widths
of X-ray flares can be reproduced. In particular, the observed
$\delta t/t \sim 0.1$ trend suggests that the later flares (larger $t$)
are wider (larger $\delta t$). This requires that the shell width
$\Delta$ broadens with time. A natural broadening mechanism is
shell spreading (Eq.[\ref{shellspreading}]). After a shell enters
the spreading regime, the width of the shell is proportional to
the radius, so that resultant X-ray flare width can be wide if the
collision radius is large. Without placing the blast  wave constraint,
one indeed expects many ``fat'' flares, corresponding to very large
radius collisions. With the blast wave constraint, the number of
``fat'' flares reduces significantly. Figure \ref{widthinfo}a displays
the model-predicted flare width as a function of their occurrences
for a narrow distribution of the initial width $\Delta_0$ (between
$10^{10}-2\times 10^{10}$ cm,
with the $\delta t/t=0.1$ line over plotted.
 The blue pluses are
allowed but the red crosses are disallowed. The result suggests
that although the predicted values are around the $\delta t/t=0.1$
line, the scatter is broader than what is observed. In particular,
many narrow flares are predicted (those without significant
spreading). The lightcurve for this simulation is presented in Fig.\ref{latefatshells}a,
which shows narrow late time flares that are not observed by the
Swift/XRT data. In order to reproduce the observations, one is
required to increase the initial shell width $\Delta_0$ for late
ejection episodes. Figures \ref{widthinfo}b and \ref{latefatshells}b show an example
of late injection of ``fat'' shells. Generally, $\Delta_0
\propto t_{ej}$ is needed to reproduce the observed data.
This is consistent with the expectations of some central engine models.
For example, in the fragmented disk model proposed by Perna et al.
(2006), the clumps at larger radii have lower densities and tend to be
more spread out so that the accretion time scale is longer. The
ejected shells correspondingly also have longer durations.

\begin{figure}
\plotone{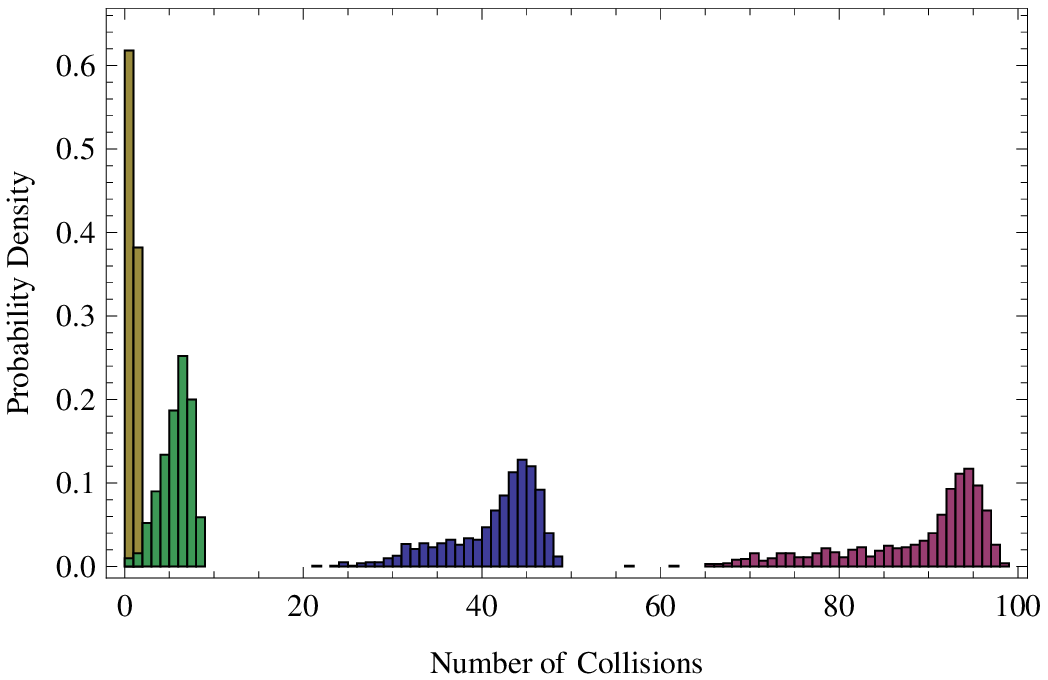}
  \caption{The most likely number of allowed collisions for
    simulations with 4, 10, 50 and 100 shells.  1000 simulations were
    run for each.}
\label{stats}
\end{figure}

\section{Conclusion}

We have developed a numerical code to model the internal collisions of
an unsteady wind with arbitrary central engine activities. This is an
extension of the internal shock models previously developed to model
GRB prompt emission (e.g. Kobayashi et al. 1997), with a focus on
X-ray flares that are commonly observed in GRB X-ray afterglows
(e.g. Chincarini et al. 2007). Our motivation is to diagnose the
required central engine activities based on the observational
properties of X-ray flares as summarized in \S1. The following conclusions
can be reached:

\begin{itemize}

\item The internal shock model with multiple ejection episodes can
generally reproduce the properties of X-ray flares. Our shell model
naturally explains both prompt emission and X-ray flares, suggesting
that they originate from very similar mechanisms.
\item We find that the number of pulses/flares is directly related to
the number of shells released from the central engine. More shells means
more flares.  In general, we find that the number of collisions is
slightly smaller than the number of the shells ejected (e.g. the most
probable values are: $5\%$ less for 100 simulated shells, $10\%$ less
for 50 simulated shells, 8 collisions for 10 simulated shells and 1
collision for 4 simulated shells, see Fig.\ref{stats} for
distributions). For a modal number of observed flares of 1, between
2 and 5 shells need to be released on average.
\item The correlation between $t_{ej}$ and $t_{\oplus,col}$ ensures
that the shells that are responsible for creating flares must be
ejected just prior to being seen. This is because the second term
in the right hand side of Eq.(\ref{time}) is typically much
shorter than the first term. This means that not only the central
engine activity must be prolonged, but it must also be episodic.
Steady energy injection cannot produce flare-like features.  In
other words, early flares are created by shells ejected early, while
late flares by shells ejected late.  Since flares are seen as late as
$10^6$ seconds, this means that the central engine can be active for
a long time after the prompt emission. These conclusions are
consistent with Zhang et al. (2006), Liang et al. (2006) and
Lazzati \& Perna (2007).
\item The large variance of fluences seen in flares can be explained
by the different energies ($E~\sim \gamma m c^2$) of the ejected
shells.
The peak luminosity of a flare also depends on the width of the pulse,
which is either related to shell spreading or intrinsically different
durations of shell ejection.
\item This study seems to rule out uniform (or narrow distribution)
thickness shells.  The observed widths of flares cannot be reproduced
solely by spreading effects. One requires that later ejection episodes
eject shells with larger widths to explain the typical flare width of
$\delta t_e/t \sim 1/10$.  Since most shells collide before spreading,
typical late collisions would be too narrow if later ejected shells
had the same width as the initial prompt ejection episodes.
Sharp flares where the width is near 1/10 of the emission time
of the flare are common, but ``fat'' flares are seen occasionally as
well.  For example, the giant flare of GRB050502B has a
$\delta t_e/t \sim 1$ (Chincarini et al. 2007) implying that shells
may either have spread before colliding or may simply have had a large
width when ejected.  Shells which have spread significantly before
colliding will have energies spread over a large log-Gaussian shape
and may be too dim to reach above the afterglow decay.

\item For goldilocks type shells, flares are seen in the XRT band
only.  The gamma-ray component of the X-ray flare (Band-function
extension)
is below the BAT detector threshold.  Figure \ref{fluxlightcurves}
shows light curves from a few typical simulations and these features
can be seen.
\item Flares superimposed on other flares are simply shells that
collide near the same time with different widths and fluences.
The observed X-ray afterglow is a superposition of flares due to
prolonged central engine activity and a background afterglow
radiation, whose origin is not addressed in our paper.

\item The decrease in average flare luminosity as a function of time
indicates shells released at later times must create less energetic
collisions (Lazzati et al. 2008). In order to produce the results
seen, late released shells must be wider, less dense, slower or a
combination of the above in order to create less energetic collisions
seen.

\item The same shell model can give a prediction on the X-ray
afterglow emission from the blast wave. Due to the continuous piling
up of late-ejection shells onto the blast wave, the lightcurve can
show a shallow decay phase that is commonly observed. However, if
a standard value $\epsilon_e=0.1$ is adopted, the predicted X-ray
lightcurve is much brighter than what is seen, by about three orders
of magnitudes. In order to reproduce the observed data, either a much
lower $\epsilon_e$ (as low as $10^{-3}$) needs to be introduced, or
the internal emission that powers the prompt emission and X-ray flares
has to be much more efficient than internal shock predictions.
\end{itemize}

Recently there has been interest in optical flares
(Kr\"{u}hler et al. 2009, Melandri et al. 2006). These flares can be
naturally interpreted in our model by invoking collisions between
low energy shells or wide shells. Such collisions could be seen in
optical bands but may be missed in higher energy bands. A direct
expectation from this model is that optical flares should on average
have lower energies and broader profiles than X-ray flares. An
internal shock origin of optical flares was also proposed by
Wei (2007).

Finally, we want to emphasize that
observationally flares have been seen in both Type I
(e.g. GRB 050724) and type II GRBs. This requires that both types
of progenitor have a similar central engine, which can eject an
episodic wind to power late central engine activities which is
a requirement of any central engine models for X-ray flares.

\acknowledgments
We thank Shiho Kobayashi and Dave Burrows for discussion, and the
anonymous referee for helpful comments.
This work is supported by NASA under grants NNX08AN24G,
NNX08AE57A, AR9-0006X, and by NSF under grant AST-0908362.

\end{document}